\documentclass[a4paper]{jpconf}
\usepackage{graphicx}
\begin{document}

\title{Dynamic masses for the close PG1159 binary SDSS\,J212531.92$-$010745.9}

\author{S Schuh$^{1}$,
        B Beeck$^{1}$ and
        T Nagel$^{2}$}
\address{$^1$Institute for Astrophysics,
  Georg-August-University of G\"ottingen,
  Friedrich-Hund-Platz~1,
  37077 G\"ottingen,
  Germany
}
\address{$^2$Institute for Astronomy and Astrophysics,
Kepler Center for Astro and Particle Physics,
  Eberhard-Karls-University of T\"ubingen,
  Sand~1, 
  72076 T\"ubingen, 
  Germany
}

\ead{schuh@astro.physik.uni-goettingen.de}

\begin{abstract}
The evolutionary scenarios which are commonly accepted for PG\,1159 stars are
mainly based on numerical simulations. These stellar evolution scenarios
have to be tested and calibrated with real objects with known stellar
parameters. One of the most crucial but also quite uncertain parameters is the
stellar mass. PG\,1159 stars have masses between 0.5 and 0.8 solar masses,
as derived from asteroseismic and spectroscopic determinations. Such mass
determinations are, however, themselves model-dependent. Moreover,
asteroseismologically and spectroscopically determined masses deviate
systematically for PG\,1159 stars by up to 10\,\%.
SDSS\,J212531.92-010745.9 is the first known PG\,1159 star in a close binary with a late main
sequence companion allowing a dynamical mass determination. 
The system shows flux variations with a peak-to-peak
amplitude of about 0.7 mag and a period of about 6.96\,h. In August 2007, 13
spectra of SDSS\,J212531.92-010745.9 covering the full orbital phase range were taken
at the TWIN 3.5\,m telescope at the Calar Alto Observatory (Alm\'{e}ria,
Spain). These confirm the typical PG\,1159 features seen in the SDSS discovery
spectrum, together with the Balmer series of hydrogen in emission
(plus other emission lines), interpreted as signature of 
the companion's irradiated side. A radial velocity curve was obtained for
both components. Using co-added radial-velocity-corrected spectra, the spectral
analysis of the PG\,1159 star is being refined.
The system's lightcurve, obtained during three seasons of photometry
with the G\"ottingen 50cm and T\"ubingen 80cm telescopes, was fitted
with both the \texttt{NIGHTFALL} and \texttt{PHOEBE} binary simulation programs. 
An accurate mass determination of the PG\,1159 
component from the radial velocity measurements requires to first derive
the inclination, which requires light curve modelling and 
yields further constraints on radii, effective temperature and
separation of the system's components. From the analysis of all data
available so far, we present the possible mass range for the PG\,1159
component of SDSS\,J212531.92-010745.9.
\end{abstract}
\section{The SDSS\,J212531.92$-$010745.9 system: a close PG\,1159 binary} %
\subsection{Evolutionary scenarios and masses for PG\,1159 stars}
\begin{figure}
\begin{center}
\includegraphics[width=0.85\textwidth]{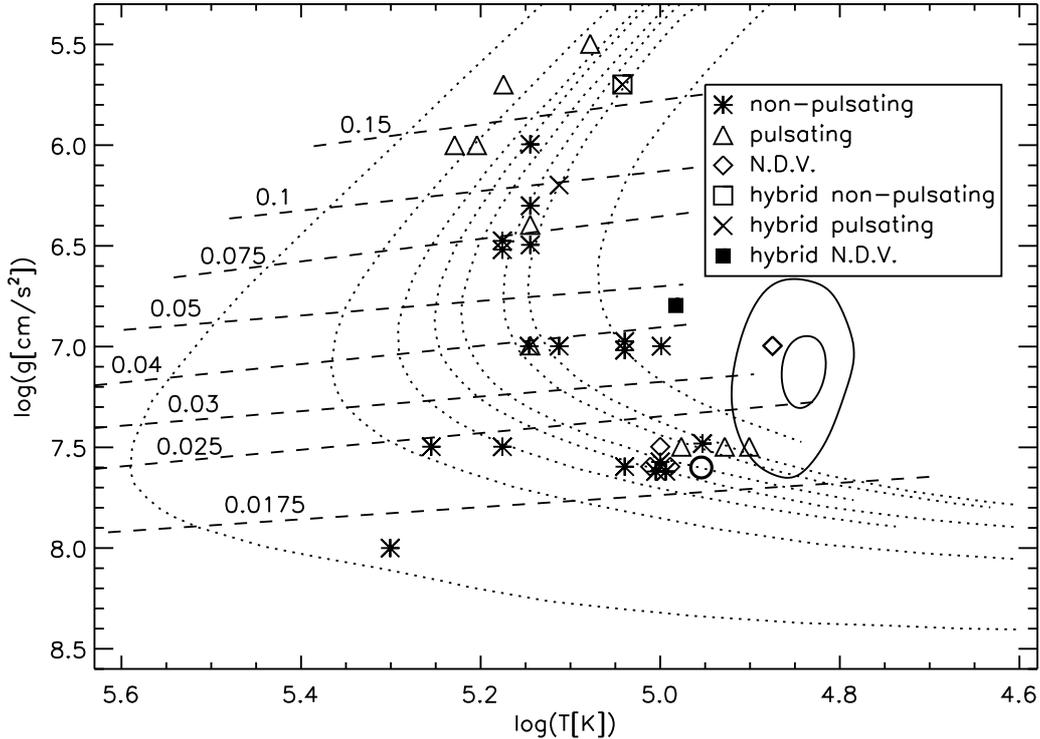}
\end{center}
\caption{\label{fig:tracks}Positions of PG\,1159 stars with
  known $T_{\rm eff}$ and $\log{g}$ in the log $T_{\rm eff}$-$\log{g}$-diagram.
  The current state-of-the-art post-AGB evolutionary tracks, first
  presented by Miller Bertolami et al.\ (2006a), as well as the
  positions of known PG1159 stars, are adopted from
  Figure~3 in Miller Bertolami \& Althaus (2006b, dotted lines).
  The masses correspond to evolutionary sequences for 0.87, 0.664,
  0.609, 0.584, 0.564, 0.542, 0.53, and 0.512~M$_\odot$ from left to right.
  We also give lines of constant radii for the theoretical tracks
  (dashed lines with annotations in R$_\odot$).
  The position of SDSS\,J212531.92$-$010745.9 found by Nagel et
  al.\ (2006) is shown as a circle; the position we find from our
  new spectral analysis (see section~\ref{sec:spectralanalysis}) is
  indicated by the error ellipse.
}
\end{figure}

PG\,1159 stars are hot hydrogen-deficient (pre-)white dwarfs with effective
temperatures between 75\,000 and 200\,000\,K, and $\log{(g/\textrm{cm}\,\textrm{s}^{-2})}$\,=\,5.5--8.0 (Werner 2001).
They are in the transition between the asymptotic giant branch (AGB) and cooling white dwarfs.
Spectra of PG\,1159 stars are dominated by absorption lines of He\,{\sc ii}, C\,{\sc iv}
and O\,{\sc vi}. Roughly half of them have a planetary nebula.
Current theory suggests (e.\,g.~Werner 2001) that they are the outcome
of a late or very late helium-shell
flash, a phenomenon that drives the currently observed fast evolutionary
rates of three well-known objects (FG~Sge, Sakurai's object, V605
Aql). Flash-induced envelope mixing produces a H-deficient stellar
surface. The photospheric composition then essentially reflects that of the
region between the H- and He-burning shells in the precursor AGB star. The
He-shell flash forces the star back onto the AGB. The subsequent, second
post-AGB evolution explains the existence of Wolf-Rayet central stars of
planetary nebulae and their successors, the PG\,1159 stars. These in turn
are the progenitors of the helium rich WD sequence. Currently, 40 PG\,1159
stars are known, Figure\,\ref{fig:tracks} shows their positions in a log\,$T_{\rm eff}$-$\log g$-diagram.
\par
Recent evolutionary models for PG\,1159 stars (Miller Bertolami \& Althaus
2006b, Werner \& Herwig 2006) lead to spectroscopic masses which still
suffer from serious uncertainties concerning the efficiency of the
overshooting process in the He-flash driven convection zone during the
thermal pulse phase on the AGB. This efficiency is represented in the
models with an a priori unknown free parameter. The other ingredient
for the determination of spectroscopic masses, the stellar atmosphere
models, have to deal with uncertainties in the line broadening theory,
so lacking reliable physical input there makes the determination of
$\log{g}$ particularly difficult. 
\par
The overall resulting uncertainties in model masses manifest themselves as 
discrepancies between current asteroseismological masses and
spectroscopic masses, as is evident from Table~3 in Werner \& Herwig
(2006; an updated version has been adopted as Table~1 in Schuh \&
Nagel 2007). The determination of a dynamical mass as an alternative method
and test case for the models is currently restricted to only one
suitable object: SDSS\,J212531.92$-$010745.9.
\begin{figure}
\begin{center}
\includegraphics[width=0.9\textwidth]{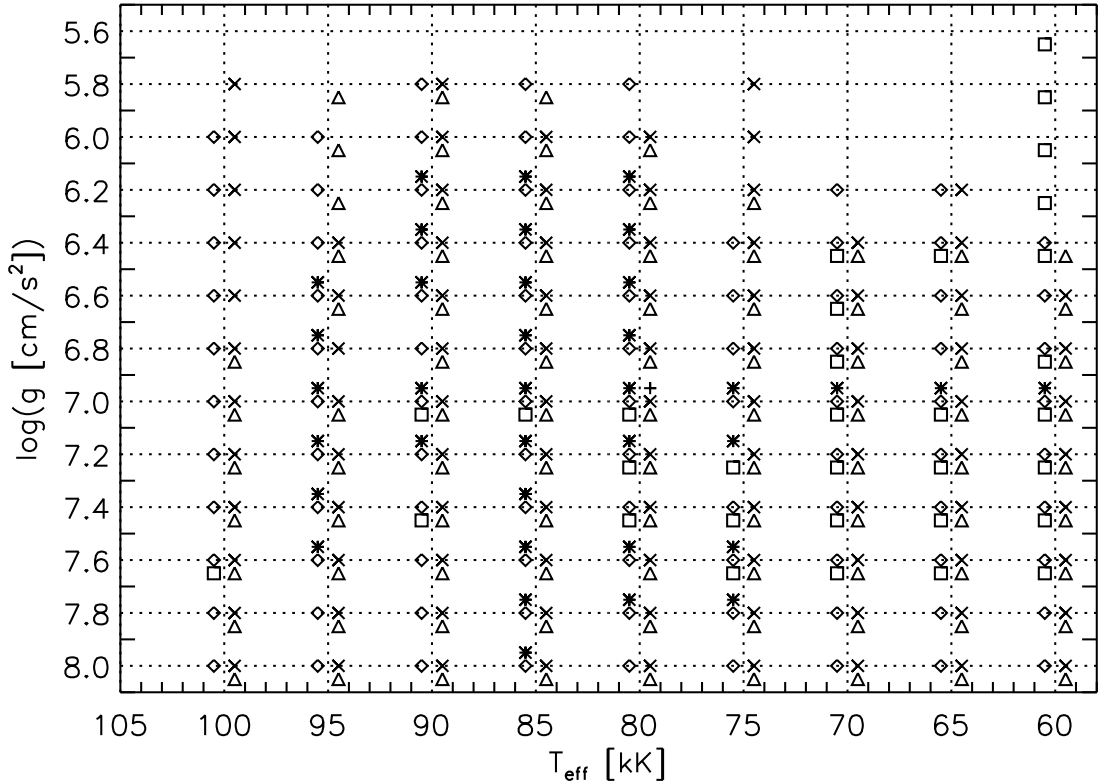}
\end{center}
\caption{\label{fig:modelgrid}The model atmospheres were calculated
  with \texttt{NGRT} (Werner et al.\ 2003). Shown are the models
  calculated by end of October 2008 in the $\log{g} -
  T_{\textrm{eff}}$-plane for six different \textsc{C}-abundances
  represented by different plot symbols (squares: \textsc{C/He}=0.01, triangles:
  \textsc{C/He}=0.03, diamonds: \textsc{C/He}=0.05, crosses: \textsc{C/He}=0.07, asterisks:
  \textsc{C/He}=0.10, plus signs: \textsc{C/He}=0.13). Other elemental
  abundances (all by number) are fixed at \textsc{N/He}=0.01 and
  \textsc{O/He}=0.01. The grid is still being completed:
  The whole range from $T_{\textrm{eff}}=60\ldots100\,\textrm{kK}$ in steps of
  5 kK and from $\log{(g/\textrm{cm}\,\textrm{s}^{-2})}=5.6\ldots8.0$ in steps of 0.2\,dex will be filled
  with models in all six \textsc{C}-abundances by December 2008. Model spectra
  were calculated from the atmosphere models using \texttt{profile} and
  were convolved with a gauss profile to match the spectral resolution
  of $1.2\,\textrm{\AA}$ of the Calar Alto TWIN Spectograph in the
  configuration used.  
}
\end{figure}

\begin{figure}
\begin{center}
\includegraphics[width=0.9\textwidth]{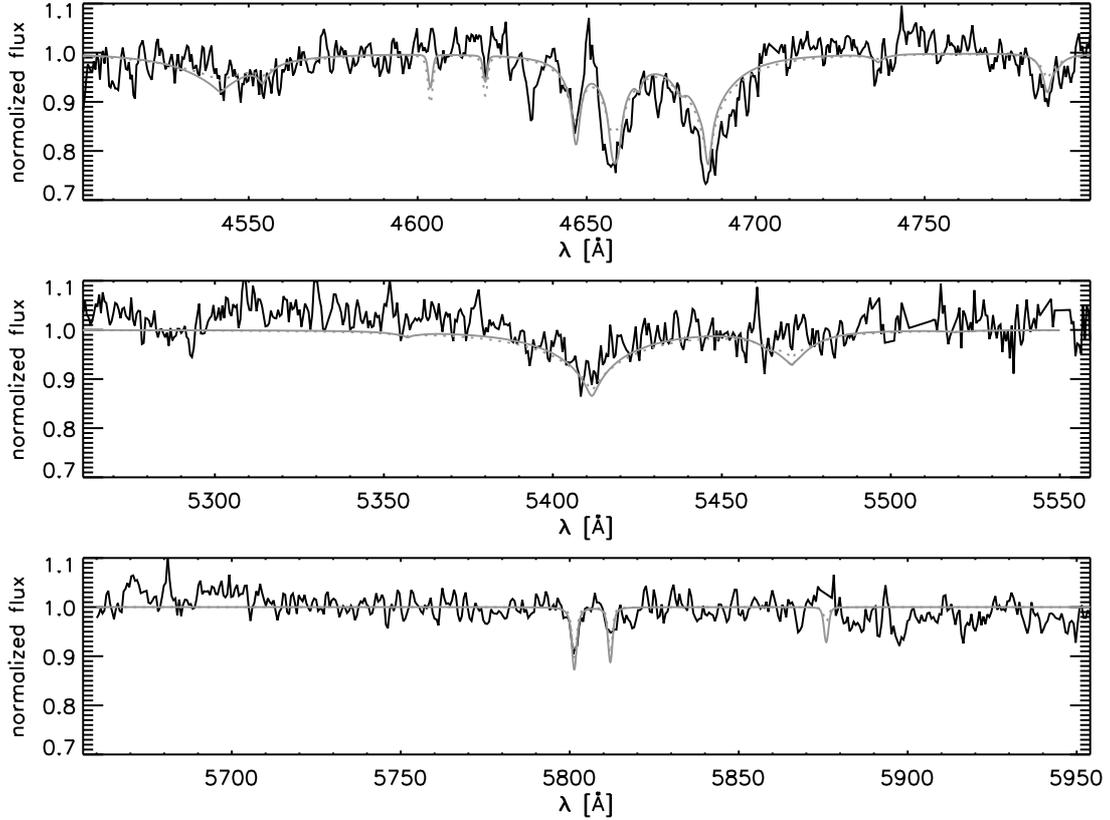}
\end{center}
\caption{\label{fig:spectralfit}
  Cross-correlating specific spectral features of the Calar Alto
  spectra radial velocities were deduced. The
  three panels show three distinctive PG1159 features in the median
  spectrum (solid black line) along with the two model spectra (grey
  lines), one of which (dotted line) belongs to the model atmosphere
  with $\log{(g/\textrm{cm}\,\textrm{s}^{-2})} = 7.60$ and $T_{\textrm{eff}}=90000\,\textrm{K}$ and
  $\textsc{C/He}=0.05$ and was found as first preliminary result by
  Nagel et al.\ (2006). The solid grey line is the model spectrum with
  the highest cross-correlation value found by the quantitative spectral
  analysis with the extended model grid ($\log{(g/\textrm{cm}\,\textrm{s}^{-2})} = 7.1$,
  $T_{\textrm{eff}}=72500\,\textrm{K}$, $\textsc{C/He}=0.07$). 
  The lowermost panel shows the narrow \textsc{Civ} lines used for the radial velocity
  determination of the PG1159 star.
}
\end{figure}

\begin{figure}
\begin{center}
\includegraphics[width=0.86\textwidth]{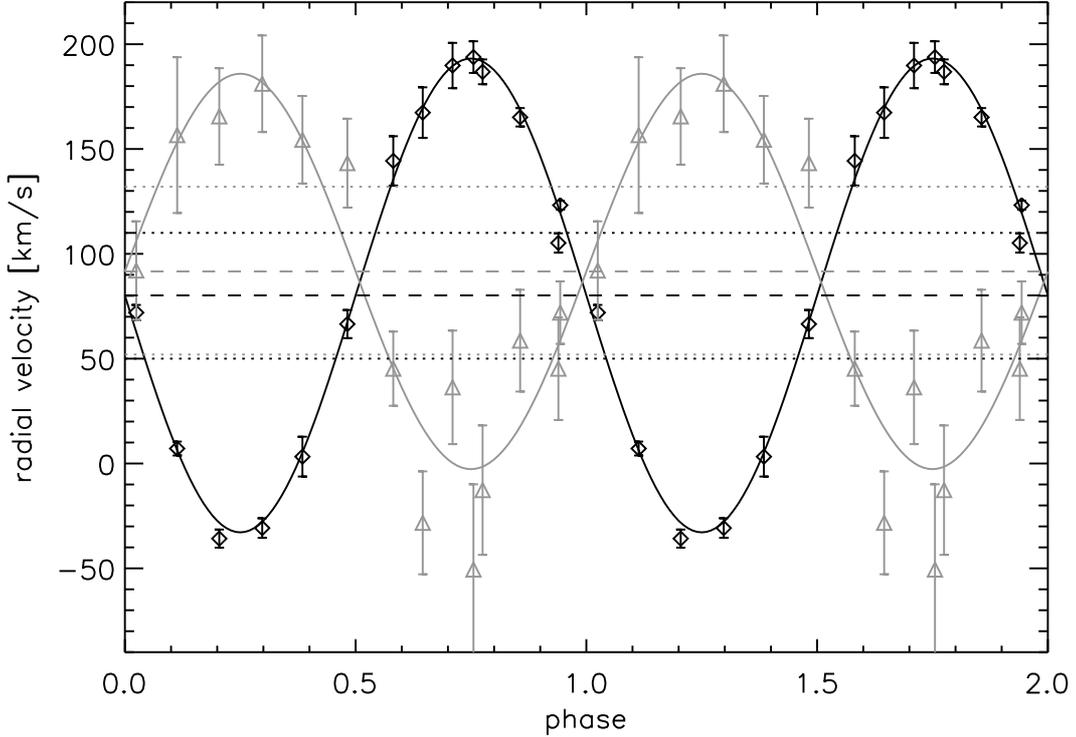}
\end{center}
\caption{\label{fig:radialvelocities}Radial velocity curves for the
  PG\,1159 star (triangles, light gray) and the companion (diamonds, black)
  determined from spectra obtained with TWIN at Calar Alto. The best fit
  sine curves yield semi-amplitudes of 
  $k_1 = 94.3 \pm 15.0\, \textrm{km}\,\textrm{s}^{-1}$ for the PG\,1159
  (light gray sine curve) and 
  $k_2 = 113.0 \pm 3.0\, \textrm{km}\,\textrm{s}^{-1}$
  for the the companion (black sine curve). The zero points of the two
  velocity curves are shown as dashed lines, the dotted lines
  indicate their uncertainties. The shift of the zero points
  relative to each other by about 11\,km\,s$^{-1}$ could be explained by
  gravitational redshift of the PG\,1159 star but is not significant.
}
\end{figure}

\begin{figure}
\begin{center}
\includegraphics[width=0.86\textwidth]{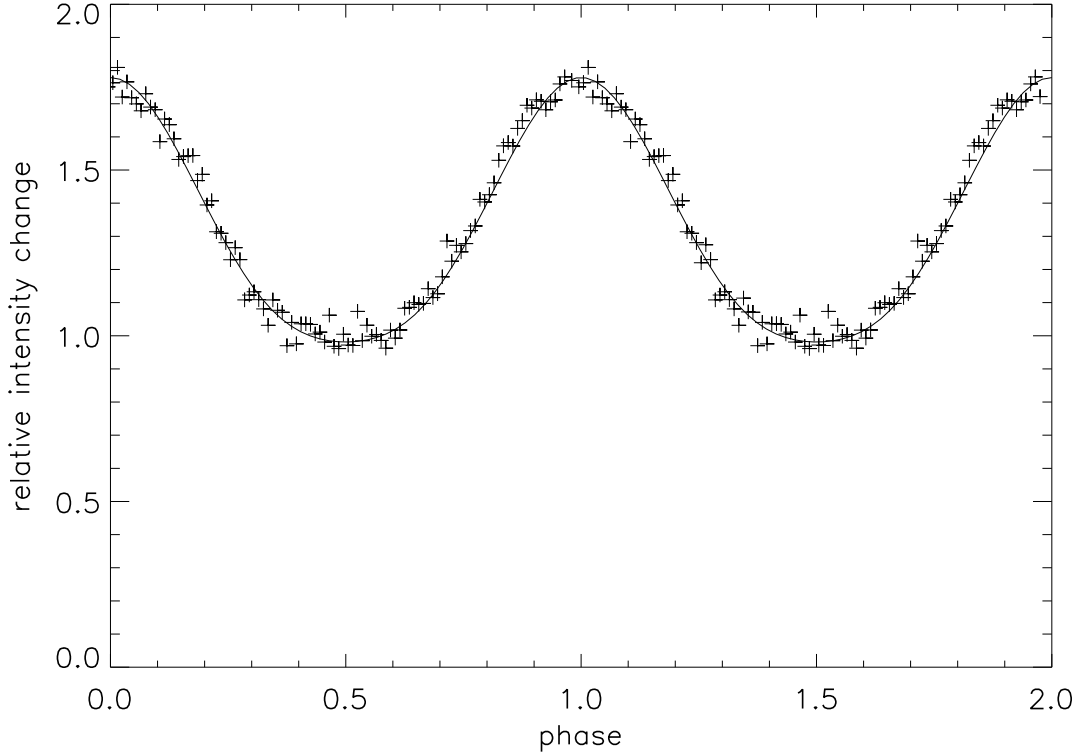}
\end{center}
\caption{\label{fig:lightcurve}The observed light curve of all nights from 2005
  to 2007 folded onto the orbital period and the
  simulated light curve of a binary system, consisting of a PG\,1159 
  star and a heated M dwarf, calculated with
  \texttt{PHOEBE} for $m_1$=0.55\,M$_\odot$, $\log{(g/\textrm{cm}\,\textrm{s}^{-2})}$=7.1,
  $R_1$=0.035\,R$_\odot$(black line).
}
\end{figure}

\begin{figure}
\begin{center}
\includegraphics[width=0.9\textwidth]{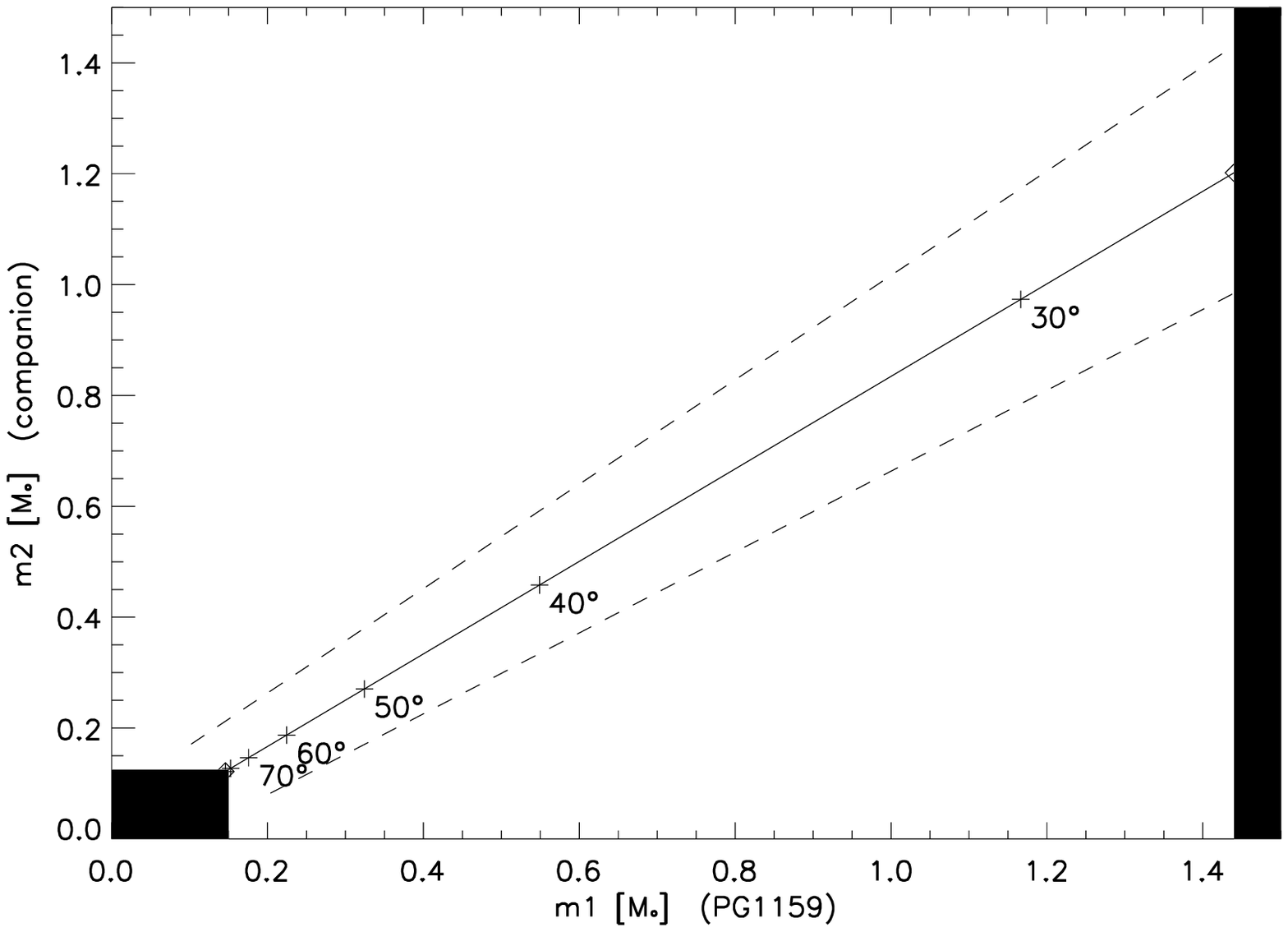}
\end{center}
\caption{\label{fig:msin3iconstraints}Using the photometric period and
the mass ratio from the radial velocity amplitudes, the masses can be
obtained as a function of the inclination. For the radial velocity
curve solution in Figure\,\ref{fig:radialvelocities}, the
($m_1$,$m_2$) pairs lie on the line (error range: dashed) shown above,
with $m_{1,\textrm{min}}$=0.15\,M$_{\odot}$ and
$m_{2,\textrm{min}}$=0.12\,M$_{\odot}$. The dark areas mark
forbidden ranges of the inclination: $m_1<1.44$\,M$_{\odot}$ requires
$i>20^{\circ}$, and the fact that the system is not eclipsing sets a
limit of $i< 80^{\circ}$, implying $m_1>0.15$\,M$_{\odot}$; more
stringent constraints are clearly required. 
A comparison with the evolutionary tracks in Figure\,\ref{fig:tracks}
(although these may not be appropriate for a presumed PCEB system) favours a
higher inclination (lower masses) than assumed in 
Figure\,\ref{fig:lightcurve}.  
}
\end{figure}

\subsection{Discovery of one PG\,1159 close binary among many SDSS white
dwarf binaries}
The important role of binaries containing white dwarfs, for dynamical
mass and other parameter determinations, is evident from the number of
publications devoted to the topic in these proceedings. The
contributions by Heller et al.\ (2009), Rebassa-Mansergas et al.\
(2009), Schreiber et al.\ (2009), and further contributions on 
specific individual systems, reflect the remarkable increase in number
of known white dwarfs with companions that the Sloan Digital Sky
Survey (SDSS) has brought about.
Among the PG\,1159 stars, however, SDSS\,J212531.92$-$010745.9 is the
only known close binary system (Nagel et al.\,2006). Eisenstein et
al.\ (2006) also list it as a spectroscopically confirmed white dwarf
from the SDSS. 
\par
The spectrum of SDSS\,J212531.92$-$010745.9 shows significant
features that are typical for PG\,1159 stars, for example the strong
C\,{\sc iv} absorption lines at $4650-4700$\AA\ and He\,{\sc ii} at $4686$\AA.
Furthermore, the spectrum shows the Balmer series of hydrogen in
emission. We have shown from our extensive follow-up photometry and
modelling that this is due to a close cool companion which is heated up by
irradiation from the hydrogen-deficient PG\,1159 star. Thus, the
radial velocities for both stars can be measured in this double-lined
system.
\par
Nagel et al.\ (2006) first reported a photometric period of
6.95573(5)\,h with an amplitude of 0.354(3)\,mag which represents the
orbital period of the binary system.  From an initial comparison of
the SDSS archive spectrum (Data Release~4) with NLTE model
spectra they derived, as preliminary 
results, an effective temperature of 90\,000\,K,
$\log{(g/\textrm{cm}\,\textrm{s}^{-2})}\,\sim$\,7.60 and the abundance ratio
\textsc{C/He}\,$\sim$\,0.05 for the PG\,1159 component. They simulated the
light curve of the binary system using \texttt{NIGHTFALL}, showing
that in in addition to the constant contribution from the PG\,1159,
the irradiated side of the companion leads to a periodic brightening
due to the reflection effect.
For the inclination they obtained $(70\pm5)\,^{\circ}$.  The system is
not eclipsing and has not been found to be pulsating.
While their solution for a mean radius of $(0.4\pm0.1)\,\rm
R_{\odot}$, a mass of \,$(0.4\pm0.1)\,\rm M_{\odot}$ and a temperature of the
irradiated surface of about 8\,200\,K for the companion was in good
agreement with the observed light curve, a manifold of solutions
remains equally plausible as long as we cannot reliably and
indepedently fix the PG\,1159 mass and radius.
\par
We note that in the SDSS Data Release~6 catalogue Adelman-McCarthy et al.\ (2008) quote a 
proper motion of +5(3)\,mas\,yr$^{-1}$ in RA and +4(3)\,mas\,yr$^{-1}$
in DEC, a value for the redshift corresponding 
to 138(60)\,km\,s$^{-1}$, and Sloan colours of
u'=17.144(0.009)\,mag,
g'=17.538(0.005)\,mag,
r'=17.757(0.006)\,mag,
i'=17.794(0.008)\,mag,
z'=17.828(0.020)\,mag.
\subsection{Orbit ephemeris from photometric observations}
From time-series photometry of SDSS\,J212531.92$-$010745.9 performed
in 2005, 2006 and 2007 with the T\"ubingen 80\,cm and the G\"ottingen
50\,cm telescopes, Schuh et al.\ (2008b) find the ephemeris of the predicted
\emph{maxima} times to be ~
\mbox{$
{\rm HJD} = 2454055\hbox{$.\!\!^{\rm d}$} 2134(4)  +    0\hbox{$.\!\!^{\rm d}$} 289822(2) \cdot E.
$}
Further multi-colour observations are available, which show that the
photometric amplitude in the optical is roughly 35\% smaller in the
blue than in the red.
\section{Analysis of the Calar Alto 3.5\,m TWIN spectroscopic data}
\subsection{Observational data and stellar atmosphere model grid}
In 2007 we obtained time-resolved spectra with the TWIN spectrograph
at the 3.5\,m telescope at Calar Alto. The observations have been described
in Schuh et al.\ (2008a): in the double-lined spectrum
both sets of lines vary as expected with the orbital phase. The series of
phase-resolved spectra can be used to derive radial velocity curves on
the one hand, and to refine the spectral analysis (originally done on
one SDSS spectrum) on the other hand.
In order to do a spectral analysis of the PG\,1159 component, a
suitable grid of stellar model atmospheres is being calculated using
\texttt{NGRT} (see Figure\,\ref{fig:modelgrid} for details).
\subsection{Spectral analysis of the PG\,1159 component}
\label{sec:spectralanalysis}
The spectra were wavelength shifted according to the radial velocity
solution shown in Figure\,\ref{fig:radialvelocities} and co-added,
however without correcting for the phase-variable continuum
contribution by the irradiated side of the companion yet, for a
spectral analysis with our model grid. 
The median spectrum of the PG\,1159 component was
normalised to the continuum and cross-correlated with each normalised
spectrum of the model grid. This cross-correlation was 
restricted to wavelength ranges where at least one of the model spectra shows
significant spectral features (minimum line depth is a free parameter of this
restriction; usually 1\% was chosen, 3\%, however, yields similar
results) and where the mean companion spectrum shows no significant spectral
features (the maximum line height of emission lines is another free parameter for the
restriction -- possible red shift differences between the two components are taken into
account). Different choices of the free parameters show a stable maximum
in the cross-correlation values for the median spectrum at an effective
temperature of about $(72500\pm5000)\,\textrm{K}$, $\log{(g/\textrm{cm}\,\textrm{s}^{-2})}$ of about $7.1\pm
0.5$ and \textsc{C/He} of $0.07\pm0.03$. The error bars quoted are very
preliminary by-eye values.
Two spectra from the model grid are shown in comparison to three
spectral regions containing lines of the
PG\,1159 component of SDSS\,J212531.92$-$010745.9 in
Figure\,\ref{fig:spectralfit}.
Combing the quantitative with the by-eye spectral
analysis, we currently constrain the stellar parameters of the PG\,1159
component to the following cuboid in parameter space:
$T_{\textrm{eff}}=65000K\ldots90000\,\textrm{K}$, $\log{(g/\textrm{cm}\,\textrm{s}^{-2})} =6.50\ldots8.00$,
$\textsc{C/He}=0.04\ldots0.10$. These preliminary values were obtained
as part of the diploma thesis work by Beeck (2009) and will
be finalised when the model grid has been completed and the
quantitative spectral analysis has been optimised.
\subsection{Radial velocity curves}
The radial velocity curves derived by Beeck (2009) are shown
in Figure\,\ref{fig:radialvelocities}.
Due to the individual exposure times of 30\,min the
spectral lines are somewhat smeared by the orbital motion, adding to the
uncertainties in the determination of the radial velocities.
The ratio of projected semi-amplitudes yields
$k_1/k_2\approx m_1/m_2\approx1.2$, allowing a range of mass ratios $\approx1.0\ldots1.5$.
Although we find a shift of the zero point of the PG\,1159's velocity curve with
regard to the companion's velocity curve zero point of the order of
11\,km\,s$^{-1}$ (Figure\,\ref{fig:radialvelocities}) which could be
caused by the gravitational redshift of the PG\,1159 star, this result
is not formally significant. The expected gravitational redshift for a
star with a mass of 0.5\,-\,0.8\,M$_\odot$  and a radius of
0.02\,-\,0.15\,R$_\odot$ would be of the order of 4 to 25
km\,s$^{-1}$. Better data would be required to deduce reliable
constraints for the combination of mass and radius
of the PG\,1159 star from a  gravitational redshift measurement.
\section{Phase resolved light curve and spectral modelling}
\subsection{Light curve modelling with \texttt{NIGHTFALL} and \texttt{PHOEBE}}
With the know orbital period $P$ and projected radial velocities 
$k_1$, $k_2$ measured from spectroscopy, in principle the only value
the light curve modelling still needs to provide is the inclination
$i$ to obtain the mass: \quad
\mbox{
$m_1 \sin^3{i} ~~ (k_1, k_2, P)
= \frac{P}{2\pi G} \frac{(k_1+k_2)^3}{1+\frac{k_1}{k_2}}$
}
\quad (similarly for $m_2$).
\par
At an assumed value for the inclination of 40$^{\circ}$, the
photometric variation can be fit using \texttt{PHOEBE} as shown in
Figure\,\ref{fig:lightcurve}; further independent considerations to
constrain the inclination are not overwhelmingly helpful
(Figure\,\ref{fig:msin3iconstraints}). The best value will be
determined from a comprehensive parameter study, including
a comparison of results from \texttt{NIGHTFALL} and \texttt{PHOEBE}
(Beeck 2009).
\subsection{Challenges in the modelling approach and additional constraints}
All the light curve solutions are obtained under the 
simplifying assumptions of bound rotation and a 
circular orbit. Further complications arise from the choice of the albedo value
in particular for the companion which is unknown and possibly larger
than 1 in the relevant wavelength range due to redistribution, 
large uncertainties in the physical modelling of reflection
effect, and a formal limitation inherent to \texttt{PHOEBE} where the 
maximum $T_{\textrm{eff}}$ currently available is 50\,000\,K.
\par
Constraints for the many parameters governing a light curve simulation
will come from the spectral analysis not only of the PG\,1159, but
also from irradiated spectra calulated using \texttt{PHOENIX} to model
the hot side of the companion.
A particularly strong additional constraint would come from a good
determination of the gravitational redshift for the PG\,1159 star.
\ack{
We would like to thank B.~G\"ansicke and M.~Schreiber for
pointing out to us the interesting SDSS spectrum of this object.
We also thank S.~Dreizler and S.~H\"ugelmeyer
for their help in setting up the stellar atmosphere model
calculations, and K.~Werner and M.~Miller-Bertolami for inspiring
discussions. Thanks to I.~Traulsen and many other observers for
supporting the photometric observations in G\"ottingen and T\"ubingen.
The spectroscopy is based on observations collected in service
mode by Javier Aceituno and Ulrich Thiele for our program
\mbox{H07-3.5-038} at the Centro Astron{\'o}mico Hispano
Alem{\'a}n (CAHA) at Calar Alto, operated jointly by the Max-Planck
Institut f\"ur Astronomie and the Instituto de Astrof{\'i}sica de
Andaluc{\'i}a (CSIC).
Travel to the 16th White dwarf workshop where this
contribution was presented has been subsidised by the Deutsche
Forschungsgemeinschaft (DFG) under grant number \mbox{KON 856/2008
  SCHU 2249/4-1}. 
}

\section*{References}

\begin{thereferences}

\item 
Adelman-McCarthy J~K et al.\ 2008, {\sl APJS} {\bf 175} 297 

\item
Beeck B 2009, University of G\"ottingen,
{\sl diploma thesis}, in prep.

\item 
Eisenstein D J et al.\ 2006, 
{\sl APJS} {\bf 167} 40 

\item 
{Heller R et al.\ 2009, {\sl these proceedings} }

\item 
Miller Bertolami M~M, Althaus L~G, Serenelli A~M and Panei J~A 2006a, 
{\sl A\&A} {\bf 449} 313 

\item 
Miller Bertolami M~M and Althaus L~G 2006b, {\sl A\&A} {\bf 454} 845 

\item 
Nagel T, Schuh S, Kusterer D-J, Stahn T, H{\"u}gelmeyer
S~D, Dreizler S, G{\"a}nsicke B~T and Schreiber M~R 2006,
{\sl A\&A} {\bf 448} L25  

\item 
{Rebassa-Mansergas A et al.\ 2009, {\sl these proceedings} }

\item 
{Schreiber M R et al.\ 2009, {\sl these proceedings} }

\item 
Schuh S and Nagel T 2007, 15th European Workshop on White Dwarfs 
{\sl ASPC} {\bf 372} 491 
 
\item 
Schuh S, Nagel T, Traulsen I and Beeck B 2008a,
{\sl Hydrogen-Deficient Stars} {\bf 391} 133  

\item 
Schuh S, Traulsen I, Nagel T, Reiff E, Homeier D, Schwager H,
Kusterer D-J, Lutz R and Schreiber M R 2008b, 
{\sl Astronomische Nachrichten} {\bf 329} 376 

\item 
Werner K 2001, {\sl Ap\&SS} {\bf 275} 27

\item 
Werner K, Deetjen, J~L, Dreizler S, Rauch T, Barstow M~A
and Kruk, J~W 2003,  {\sl White Dwarfs} {\sl NATO ASIB Proc.} {\bf 105} 117 

\item 
Werner K and Herwig F 2006, 
{\sl PASP} {\bf 118} 183 

\end{thereferences}

\end{document}